\documentclass[aps,pra,twocolumn,superscriptaddress]{revtex4}

\usepackage{amsmath,amssymb,graphicx,xcolor,bm,soul}
\usepackage{comment}
\usepackage{upgreek}
\usepackage{xfrac}
\usepackage{braket}

\definecolor{orange}{RGB}{235, 129, 0}
\usepackage[bookmarks=true,
            bookmarksnumbered=false,
            bookmarksopen=true,
            breaklinks=true,
            pdfborder={0 0 0},
            final=true,
            backref=none,
            colorlinks=true,
            linkcolor=blue,
            menucolor=blue,
            citecolor=blue,
            urlcolor=blue]{hyperref}

\begin{document}

\title{Hamiltonian operator approximation for energy measurement and ground state preparation}

\author{Tatiana A. Bespalova}
\affiliation{ITMO University, St. Petersburg, 197101, Russia}

\author{Oleksandr Kyriienko}
\affiliation{Department of Physics and Astronomy, University of Exeter, Stocker Road, Exeter EX4 4QL, UK}

\date{\today}

\begin{abstract}
The Hamiltonian operator plays a central role in quantum theory being a generator of unitary quantum dynamics. Its expectation value describes the energy of a quantum system. Typically being a non-unitary operator, the action of the Hamiltonian is either encoded using complex ancilla-based circuits, or implemented effectively as a sum of Pauli string terms. Here, we show how to approximate the Hamiltonian operator as a sum of propagators using a differential representation. The proposed approach, named Hamiltonian operator approximation (HOA), is designed to benefit analog quantum simulators, where one has direct access to simulation of quantum dynamics, but measuring separate circuits is not possible. We describe how to use this strategy in the hybrid quantum-classical workflow for performing energy measurements. Benchmarking the measurement scheme, we discuss the relevance of the discretization step size, stencil order, number of shots, and noise. We also use HOA to prepare ground states of complex material science models with direct iteration and quantum filter diagonalization, finding the lowest energy for the 12-qubit Hamiltonian of hydrogen chain H$_6$ with $10^{-5}$ Hartree precision using $11$ time-evolved reference states. The approach is compared to the variational quantum eigensolver, proving HOA beneficial for systems at increasing size corresponding to noisy large scale quantum devices. We find that for Heisenberg model with twelve or more spins our approach may outperform variational methods, both in terms of the gate depth and the total number of measurements. 
\end{abstract}

\maketitle

\section{Introduction}

Quantum computing is moving forward, and as evidenced by the quantum supremacy experiment \cite{Arute2019}, it is approaching the scale where high-performance classical computing may be challenged. However, despite ever-increasing complexity for material science simulations \cite{Kandala2019,Ganzhorn2018,IonQ2019,Arute2020}, reaching quantum advantage requires both well-tailored problems and modified quantum protocols to enable efficient computation.
While to date many quantum algorithms with favorable scaling were proposed, they typically require deep gate sequences~\cite{NielsenChuang}. Considered suitable for the future fault-tolerant devices, their realization may be years (or even decades) away. Present day error-prone devices of increased size, corresponding to noisy intermediate scale quantum (NISQ) computers \cite{Preskill2018}, require different strategies to solve a state preparation problem~\cite{BhartiRev}. One solution corresponds to the adoption of hybrid quantum-classical (HQC) approach \cite{Moll2018,CerezoRev}, where shallow depth circuits are used at the expense of increased sampling demand. The prominent example of HQC protocol is the variational quantum eigensolver (VQE) \cite{Peruzzo2014,OMalley2016,Kandala2017,McClean2016} that triggered the search for a wide range of variational quantum algorithms. HQC approaches are considered as a viable near-term strategy in quantum computing for chemistry~\cite{OxfordRev,ZapataRev}, optimization \cite{Farhi2014,GoogleQAOA,Akshay2020}, and quantum machine learning \cite{Benedetti2019}.  However, while operating at the reduced circuit depth, variational quantum algorithms also bear significant challenges, including: 1) vanishing gradients when optimizing deep quantum circuits \cite{McClean2018,Cerezo2020,Cerezo2020b}; 2) need for a suitable ansatz capable of preparing a solution state \cite{Herasymenko2019}; and 3) drastic increase of the number of terms to measure when considering the Hamiltonian averaging technique \cite{Bonet2020}. Numerous recent improvements include adaptive strategies for the ansatz search \cite{Grimsley2019,Tang2019} and automated ansatz optimization \cite{rotoselect,EVQE,Chivilikhin2020}, symmetry-preserving VQE \cite{Gard2019,Bonet2018}, natural gradient optimization \cite{McArdle2018,Stokes2020,Balint2019}, measurement frugal VQE \cite{Izmaylov2019,Kubler2020,Verteletskyi2020,Izmaylov2020,Tzu-Ching2020,Gokhale2019,Huggins2019b,Crawford2019,Cotler2020,TzuIzmaylov2020}, quantum subspace search \cite{Nakanishi2019}, layerwise learning \cite{Skolik2020}, and many others. Variational protocols were also applied to simulate the dynamics, showing promise for strongly correlated systems \cite{Li2017,Yuan2019,Chen2020,Endo2020,Endo2021}. However, the general limitation of variational approaches also calls for alternative solutions for systems at increasing size.

A different strategy to the state preparation relies on applying a non-unitary operation to cool down effectively the higher energy states. This can be done by representing them as a linear combination of unitaries (LCU) \cite{Long2008}. Being a valuable technique used in large-depth protocols for state-of-the-art Hamiltonian dynamics simulation \cite{Berry2015} and linear system of equations \cite{Childs2017}, the value of LCU approaches was also assessed for near-term devices and analog simulators where unitary evolution is an available resource. In this way by measuring wavefunction overlaps one can perform inverse quantum iteration \cite{Kyriienko2020}, quantum Krylov iteration \cite{Stair2020}, and quantum filter diagonalization \cite{Parrish2019} to study low energy properties of correlated materials and molecules. This approach was connected to the quantum version of time grid methods, that were used for the Schr\"odinger equation in the past \cite{Balint-Kurti1992}. From the variational protocols perspective, the LCU ansatz was also applied to chemistry \cite{Huggins2019} and linear algebra problems \cite{Rebentrost2019}, notably showing the ability to avoid vanishing gradient regions. 

While overall increasing requirements for the circuit depth and qubit overhead to perform SWAP test \cite{Ekert2002}, the quantum time grid methods provide a way to reach the global energy minimum when starting from a suitable initial state. As these approaches rely on simulation of unitary dynamics for the system, they can be seen as relatives of the quantum phase estimation algorithm (QPEA)~\cite{Kitaev1996,Abrams1999,Aspuru-Guzik2005,Dobsicek2007}. However, quantum time grid methods use the classical post-processing of overlaps and unlike QPEA do not require implementation of the \emph{controlled} unitary dynamics. This makes them suitable for noisy large scale quantum (NLSQ) devices, where doubling the system size is a less pertinent problem than challenges of the variational search. Time grid methods are particularly favorable for analog quantum simulators \cite{Bloch2012,Mazurenko2017} where one has the access to the simulation of quantum dynamics and overlap measurement~\cite{Islam2015,Cotler2019}. Importantly, in this case variational approaches and Hamiltonian averaging are not applicable, and quantum time grid approaches offer \emph{the} way to study low energy physics.

In this paper we propose a method to approximate a Hamiltonian operator with a sparse sum of unitary propagators. This serves as a building block for many LCU protocols, and is required when measuring the energy of the system in variational approaches, performing direct Hamiltonian iteration \cite{Ge2019}, implementing Lanczos algorithms, measuring the density of states \cite{Weisse2006} etc. 
We use a distinct differential decomposition approach, where Hamiltonian action is simulated with term-by-term quantum evolution, and the resulting energy expectation is read out as a sum of normalized overlaps. This strategy favors analog quantum simulation devices, where energy measurement through string-based Hamiltonian averaging is typically inaccessible. Comparing the proposed approach to the variational quantum eigensolver for the spin-1/2 Heisenberg model at increasing system sizes, we find that Hamiltonian operator approximation combined with quantum filter diagonalization offers better performance for lattices with more than 12 qubits, suggesting that the strategy can be used for ground state search in cases where variational methods suffer from the vanishing gradient problem.


\section{Methodology}

Our goal is to represent a Hamiltonian operator $\hat{\mathcal{H}}$ as a sum of unitary operators, taking as few terms as possible. Typically, this is done by partitioning the Hamiltonian into groups of qubit string operators (mutually commuting inside each group), as described in Hamiltonian averaging technique used to estimate energy of the system \cite{McClean2016,McClean2014}. However, certain Hamiltonians, for instance in quantum chemistry, have unfavourable scaling for the number of groups (partitions) to be measured, as it grows like $\sim O(N^4)$ ($N$ is a number of qubits) for vanilla Hamiltonian averaging \cite{Bonet2020}. Instead, we use the differential representation based on the quantum evolution operator (propagator) denoted as $\hat{\mathcal{U}}(t) = \exp(-i \hat{\mathcal{H}} t)$, where $t$ is a parameter which defines evolution time, $\hat{\mathcal{H}}$ is a Hermitian time-independent operator, and we use $\hbar = 1$ throughout the paper. Acting with the time derivative operator on the propagator, we can formally write $d\hat{\mathcal{U}}/dt = (-i \hat{\mathcal{H}}) \hat{\mathcal{U}}(t)$. The Hamiltonian operator then reads
\begin{align}
\label{eq1}
    \hat{\mathcal{H}} = i \hat{\mathcal{U}}^\dagger(t_0) \frac{d}{dt} \hat{\mathcal{U}}(t) \Big|_{t \rightarrow t_0},
\end{align}
defined using its derivative at some fixed point of time $t_0$, and it is convenient to use $t_0 = 0$.
We approximate the propagator derivative using the finite difference scheme with $S$ stencil points. The accuracy of the approximation scales as $O\left(\delta t^{S-1}\right)$ \cite{DiffErr}, where $\delta t$ is the distance between neighboring points. The derivative is approximated as a sum of unitary operators
\begin{equation}
    \frac{d \hat{\mathcal{U}} (t)}{d t} \Big|_{t \rightarrow t_0} = \frac{1}{\delta t} \sum_{n=-s}^{S-s-1} q_{n}(s) \hat{\mathcal{U}}(t_0 + n \delta t) + O\left(\delta t^{S-1}\right),
    \label{eq2}
\end{equation}
where $s$ is a shifting parameter that is arbitrary in general, and we usually choose it to be $(S-1)/2$ for symmetry reasons.
To find the expansion coefficients $q_{n}(s)$ we decompose our function at stencil points $t_0 + n\delta t$ ($n = -s, -s+1, ..., S-s+1$) into Taylor series around $t_0$, $f(t_0 + n\delta t) = \sum_{j=0}^{S-1} [(n \delta t)^j/j!] d^j f(t)/dt^j|_{t\rightarrow t_0}$ \cite{Fornberg1988}. 
Forming equations for each stencil point, we keep only the first $S$ terms in the expansions. Next, we need to compose a linear combination of these equations such that the coefficient before the first derivative is equal to $1$, and coefficients before all the other terms on the right are equal to zero.
For this we solve the eigenvalue equation $\bm{M} \cdot \bm{q} = \bm{\delta}$, where the matrix $\bm{M}$ reads
\begin{align}
    \bm{M} = \begin{pmatrix}
    1 & 1 & ... & 1\\
    -s & -s+1 & ... & S-s-1\\
    (-s)^2 & (-s+1)^2 & ... & (S-s-1)^2\\
    & &  ... &  \\
    (-s)^{S-1} & (-s+1)^{S-1} & ... & (S-s-1)^{S-1}\\
    \end{pmatrix},
\end{align}
$\bm{q} = \begin{pmatrix} q_{-s}, q_{-s+1}, ..., q_{S-s-1} \end{pmatrix}^T$ is a vector of coefficients we want to find, and we set $\bm{\delta} = \begin{pmatrix} 0, 1, ..., 0 \end{pmatrix}^T$ (as the first derivative term is considered). 
By inverting the matrix we find the required coefficients $q_{n}(s)$ and by inserting Eq.~\eqref{eq2} in Eq.~\eqref{eq1} we write the $S$-stencil differential approximation for the operator $\hat{\mathcal{H}}$ as
\begin{equation}
        \hat{\mathcal{H}} \approx \frac{i}{\delta t} \sum_{n=-s}^{S-s-1} q_{n}(s) e^{- i\hat{\mathcal{H}} (n \delta t) }  =: \hat{H}_{\mathrm{diff}}(S, \delta t).
        \label{eqExpansion}
\end{equation}
We refer to this procedure as Hamiltonian operator approximation (HOA).
\begin{figure}[t]
\includegraphics[width=\linewidth]{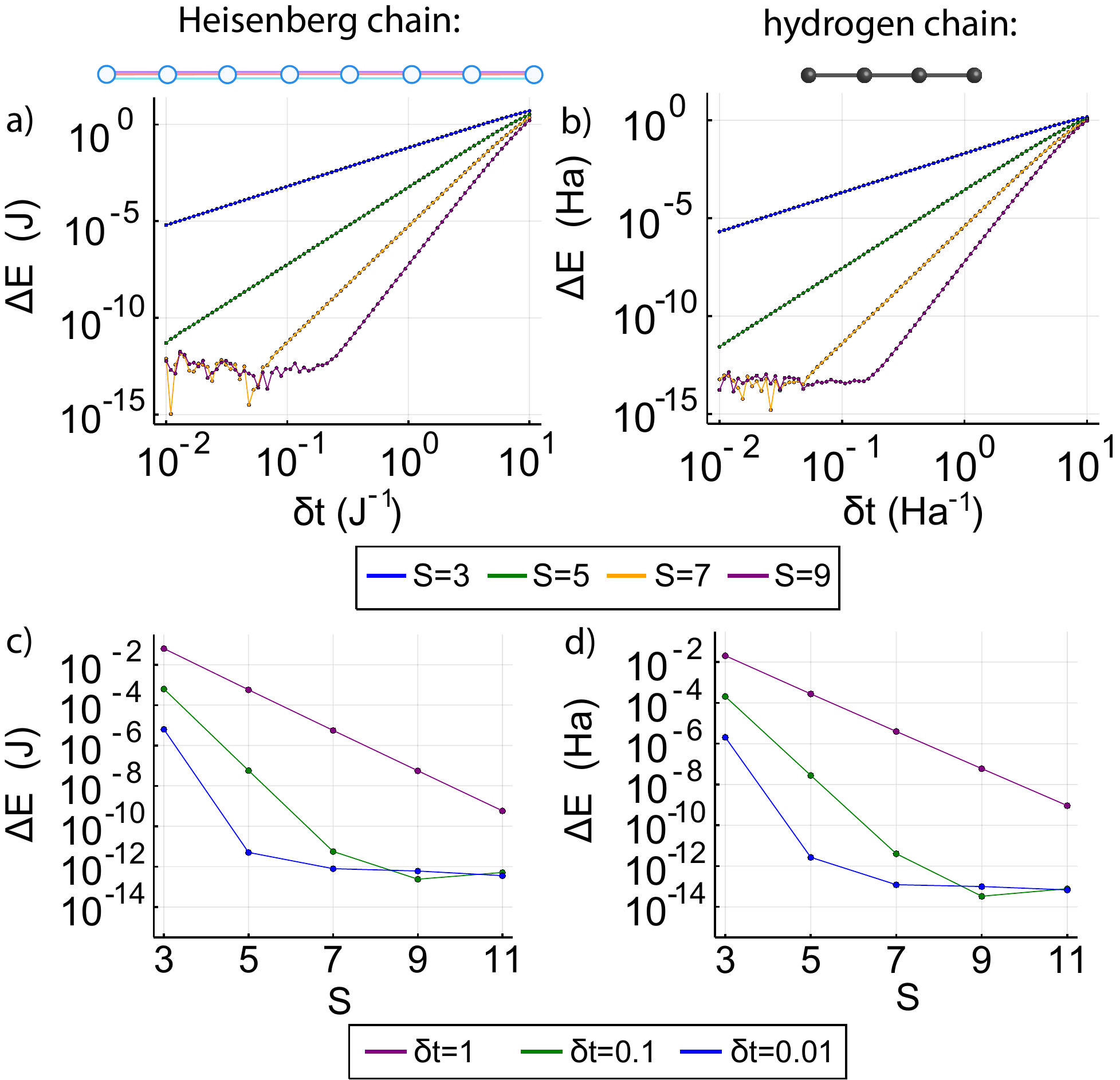}
\caption{Energy difference between exact Hamiltonian expectation and approximate energy $\Delta E$, shown for different systems pictorially presented at the top. The full statevector simulation is performed. a) $\Delta E$ as a function of time step at different stencil expansion ($S=3,5,7,9$). We consider $N=8$ Heisenberg model Hamiltonian with $h/J=0.1$, uniform state for the measurement, and plot energy in units of Heisenberg coupling $J$. b) $\Delta E$ as a function of $\delta t$ for several stencil expansions. Hamiltonian corresponds to H$_4$ hydrogen chain ($N=8$ qubits) at $d=1.0$~\AA, measured for the Hartee-Fock state. Energy is plotted in Hartree units. c) Energy difference as a function of the stencil point number $S$, plotted for $\delta t J = 1.0$, $0.1$, $0.01$. The Hamiltonian is the same as in a). d) $\Delta E$ shown for different stencil point expansions. The Hamiltonian is the same as in b).}
\label{fig:imageideal}
\end{figure}


\section{Results}

We proceed simulating complex quantum systems and study their low energy properties using the proposed Hamiltonian operator approximation approach. As test systems we choose several paradigmatic material science models. These include Heisenberg model of spin-1/2 particles in the external magnetic field, and strongly-correlated hydrogen chain as a standard example from quantum chemistry.  
For the simulation we choose the programming language {\sffamily{}Julia} and use the {\sffamily{}Yao.jl} package \cite{Luo2019} as a simulation backend, capable of performing quantum protocols with state-of-the-art efficiency~\cite{Luo2019,Zeng2019,Liu2019}.


\subsection{Energy estimation}

As one of useful applications for $\hat{\mathcal{H}}$ dynamical approximation we consider the measurement of the expected energy. Usually this is done through the procedure of Hamiltonian averaging, as commonly used in VQE \cite{McClean2016}. We propose to use a hybrid quantum-classical approach, where energy is measured as a combination of wavefunction overlaps. Using HOA [Eq.~\eqref{eqExpansion}] we can write the Hamiltonian expectation as $\langle \hat{\mathcal{H}} \rangle \approx  (i/\delta t) \sum_{n=-s}^{S-s-1} q_{n}(s) \langle \psi| \psi (n \delta t)\rangle$, where energy estimation requires calculating the overlap between $|\psi \rangle$ and the evolved state $|\psi (n \delta t)\rangle = e^{- i\hat{\mathcal{H}} (n \delta t)}|\psi\rangle$. 
Importantly, since we use the approximated derivative of the evolution operator, HOA approach favors short time evolution (see discussion below).

First, we consider the Heisenberg Hamiltonian of an $N$-qubit chain with open boundaries, $\hat{\mathcal{H}} = -J\sum_{j = 1}^{N-1} (\hat{X}_j\hat{X}_{j+1} + \hat{Y}_j\hat{Y}_{j+1} + \hat{Z}_j\hat{Z}_{j+1}) - h\sum_{j=1}^N \hat{Z}_j$, where $\hat{X}_j$, $\hat{Y}_j$, $\hat{Z}_j$ are Pauli operators acting at site $j$. 
The results of simulation are shown in Fig.~\ref{fig:imageideal}(a,c) for $N=8$ and $h/J=0.1$. We consider a uniform quantum state created by the string of Hadamard operators as $|\psi\rangle = (|0\rangle + |1\rangle)^{\otimes N}/2^{N/2}$, and measure energy expectation by evaluating terms using noiseless statevector simulation and analog unitary evolution. In Fig.~\ref{fig:imageideal}(a) we plot the difference between true expected energy and HOA expectation $\Delta E$ as a function of $\delta t$, plotted for increasing number of stencil points. We observe that the difference decreases as $O(\delta t^{S-1})$, until it reaches numerical precision-impacted region (see the full discussion on the error scaling in Appendix A). In Fig.~\ref{fig:imageideal}(c) we plot the energy difference for varying $S$ and observe exponential improvement in energy difference, which also requires smaller $\delta t$ (and correspondingly smaller circuit depth). We note that this monotonous dependence holds for the infinite number of measurement shots, and the finite shots case is discussed later.

The number of required overlaps to estimate remains small even for increasing $S$, and the complexity of HOA is defined by the maximal propagation phase $T = S \delta t$. While an analog simulation of dynamics highly benefits the proposed approach by construction, we can also use digital simulation of dynamics. Asymptotically optimal simulation of quantum dynamics can be achieved with qubitization \cite{LowChuang2019} or LCU-based approaches \cite{Berry2015}, at the expense of increased system size due to ancillary register. A resource-frugal alternative for many near- and mid-term applications can be the Trotterization approach, which was used previously for quantum time grid methods \cite{Kyriienko2020,Parrish2019}. Recent advances in Trotterized simulation of quantum dynamics suggest that scaling can be improved to $O(N T/r + N T^3 /r^2)$  for $r$ steps \cite{Tran2020}, and for local Hamiltonians there is a strong evidence that circuits can be further optimized \cite{Childs2018, Childs2019c}. For the Heisenberg model Trotterization can be readily performed by alternating couplings on even/odd sublattices \cite{LasHeras2014,Kyriienko2018}, and implemented in various hardware platforms.

To understand the scaling of HOA with Trotterized quantum evolution we study the contribution of errors coming from the product formula truncation and differentiation (see Appendix~\ref{B} for the full discussion). As HOA favors small propagation time to reduce differentiation error we observe that the number of required Trotter steps remains small for increasing system size. For $N=13$ spin-1/2 Heisenberg model, $\delta t J = 0.1$, and $S=25$, this allows for negligible Trotterization error already at $r = 5$ steps. The gate count scales as $23 N r +3N$ and gives $1534$ gates. For a hundred-qubit system with similar settings we get $\sim 10^4$ gates, which may become possible for improved NLSQ devices.

Next, we proceed studying quantum chemistry problems, and consider molecular hydrogen (open) chains H$_n$ with homogeneous bond distance $d$ between $n$ atoms. Specifically, we use examples from QunaSys competition obtained using the STO-3G minimal basis set in the spinful case~\cite{QunaSys}. Qubit-encoded Hamiltonians are then obtained using Jordan-Wigner transformation from {\sffamily{}OpenFermion} \cite{OpenFermion}.
The chain of four hydrogen atoms H$_4$ at $d = 1.0$\AA~ bond distance is encoded using 8-qubit Hamiltonian. We evaluate the energy expectation of the Hartree-Fock state, providing benchmarking in Fig.~\ref{fig:imageideal}(b,d). We find that the overall results are similar to the Heisenberg case, suggesting the internal structure of the Hamiltonian does not have a major impact on HOA accuracy.
\begin{figure}[t]
\includegraphics[width=\linewidth]{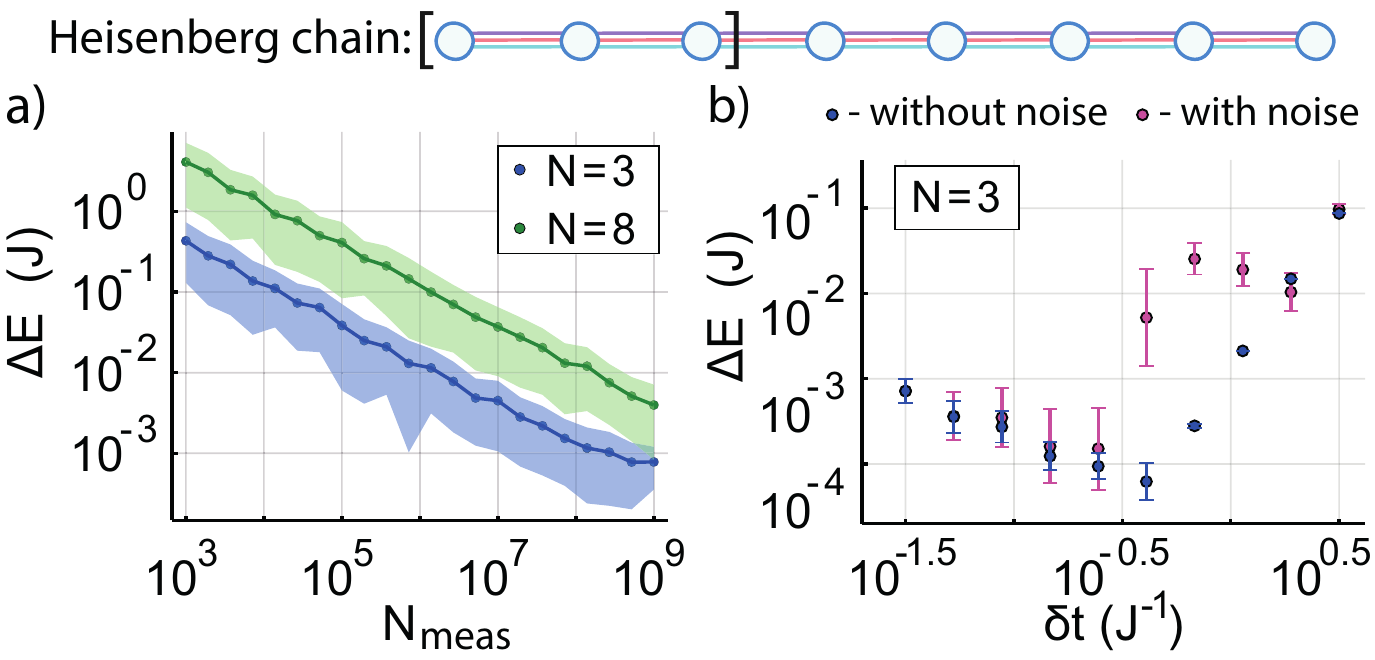} 
\caption{Influence of the number of measurements on the energy estimate via SWAP-test. a) Energy difference $\Delta E$ (in units of $J$) as a function of the number of measurements for three- and eight-qubit Heisenberg model (bottom and top lines, respectively). We set parameters to $h/J=0.1$, $S = 5$, and $\delta t J = 0.5$. To estimate the standard deviation each simulation was repeated 150 times. b) Mean energy difference as a function of $\delta t J$ for the Heisenberg model estimated taking $N_{\mathrm{meas}} = 10^9$ measurements. Blue dots correspond to noiseless simulation, with error bars showing to the standard deviation. Magenta dots and error bars are for noisy simulation performed with wave function Monte-Carlo approach for $\gamma / J = 0.1$. Other parameters are the same as in a). To estimate standard deviation each simulation was repeated 50 times.}
\label{fig:imagemeasurements}
\end{figure}


\subsection{Measurement and noise}

To assess the full performance of the Hamiltonian operator approximation, we account for other imprecision sources coming from the measurement  (finite number of shots) and effects of the environment (qubit decay).

The overlap measurement can be performed using several methods. The most popular choices correspond to SWAP and Hadamard tests \cite{Ekert2002,Higgott2019}. The SWAP test for $N$-qubit system requires additional $N+1$ qubits, where one qubit is used for the read-out \cite{Huggins2019}. A destructive version of the SWAP test can be performed using $2N$ qubits \cite{Higgott2019,Islam2015}. The Hadamard test uses a single ancilla added to the register, but doubles the circuit depth \cite{Parrish2019, Higgott2019}. We also note that in certain cases direct ancilla-free measurement is possible, for instance when circuits possess a specific structure \cite{Mitarai2019,Havlicek2019} or one of the eigenstates is known \cite{Kyriienko2020}. Further advances in the area include overlap estimation through randomized measurements \cite{Elben2020}, atom interferometry \cite{Cotler2019}, or recent projected kernel techniques \cite{Huang2021} and shadow tomography \cite{Huang2020}. The latter was proven to be optimal for predicting fidelities and entanglement entropy. We consider the improvement of overlap measurement as an important direction for advancing Hamiltonian approximation techniques.

We simulate the full measurement scheme using the SWAP test to estimate the overlap between the initial state $|\psi\rangle$ and the propagated state. 
Following the steps in \cite{Huggins2019} we prepare a superposition $\frac{1}{\sqrt{2}}\left(\ket{\psi}\ket{0} + i\hat{\mathcal{U}} \ket{\psi}\ket{1}\right)$ for the system register and an ancilla. We apply Hadamard gate to the ancilla (last qubit), and measure the expectation for $\hat{Z}_{N+1}$, getting $\langle \hat{Z}_{N+1} \rangle = -\mathrm{Im}\bra{\psi} \hat{\mathcal{U}} \ket{ \psi}$.
The results are shown in Fig.~\ref{fig:imagemeasurements} for the case of the Heisenberg model with $h/J = 0.1$, $S=5$, and $\delta t J =0.5$. We plot the difference between true energy of the uniform state in the system with Hamiltonian $\hat{\mathcal{H}}$ and the energy of the uniform state calculated with HOA, $\Delta E$, shown as a function of the number of measurement shots $N_{\mathrm{meas}}$ for which $\langle \hat{Z}_{N+1} \rangle$ is averaged. We observe that the upper bound of error decreases approximately as $\sim N_{\mathrm{meas}}^{-1/2}$ (Fig.~\ref{fig:imagemeasurements}a). Fixing $N_{\mathrm{meas}} = 10^9$ we calculate the dependence of the error on the time step $\delta t$. In the absence of noise the results are presented by blue dots in Fig.~\ref{fig:imagemeasurements}b), in line with the variance scaling $\propto c/N_{\mathrm{meas}}$ ($c$ being a system-dependent constant) found in \cite{Huggins2019}. While for relatively large $\delta t$ the error behaves similarly to ideal statevector simulator, the sampling noise increases the error at $\delta t J< 0.3$, meaning that optimal $\delta t$ depends on $N_{\mathrm{meas}}$. This suggests that resolving a small difference between two states (initial and propagated) becomes difficult at very small $\delta t$.

Next, we simulate the effect of noise using the wave function Monte-Carlo approach \cite{Dalibard1992}. For this we run $N_{\mathrm{iter}}$ trajectories of unitary evolution, subject to probabilistic action of jump operators $\sqrt{\gamma} (\hat{X}_j + i \hat{Y}_j)/2$ that can de-excite each qubit with decay rate $\gamma$. We simulate the Heisenberg model example and plot the energy deviation as a function of $\delta t$ (Fig.~\ref{fig:imagemeasurements}b, magenta bars). The effect of decay is pronounced at large times $\delta t J > 1$, where a significant number of jumps leads to erroneous overlap estimates at impacted trajectories (we consider large decay rate $\gamma/J = 0.1$, and simulate $10^9$ trajectories). As $\delta t$ and maximal propagation time $T$ decrease, so does the jump probability, leading to improved energy precision $\Delta E$. However, for small $\delta t J < 0.1$ we find that even a small number of jumps increases the estimate imprecision, given that each collapsed trajectory leading to corrupted overlap estimate enters with $\delta t^{-1}$ weight. We observe that optimal $\delta t \approx 0.2 J^{-1}$ lies close to $N_{\mathrm{meas}}$-defined case, depending on decay rate $\gamma$, and we note that a small time step region is the overall preferred choice for HOA.
\begin{figure}[t]
\includegraphics[width=0.8\linewidth]{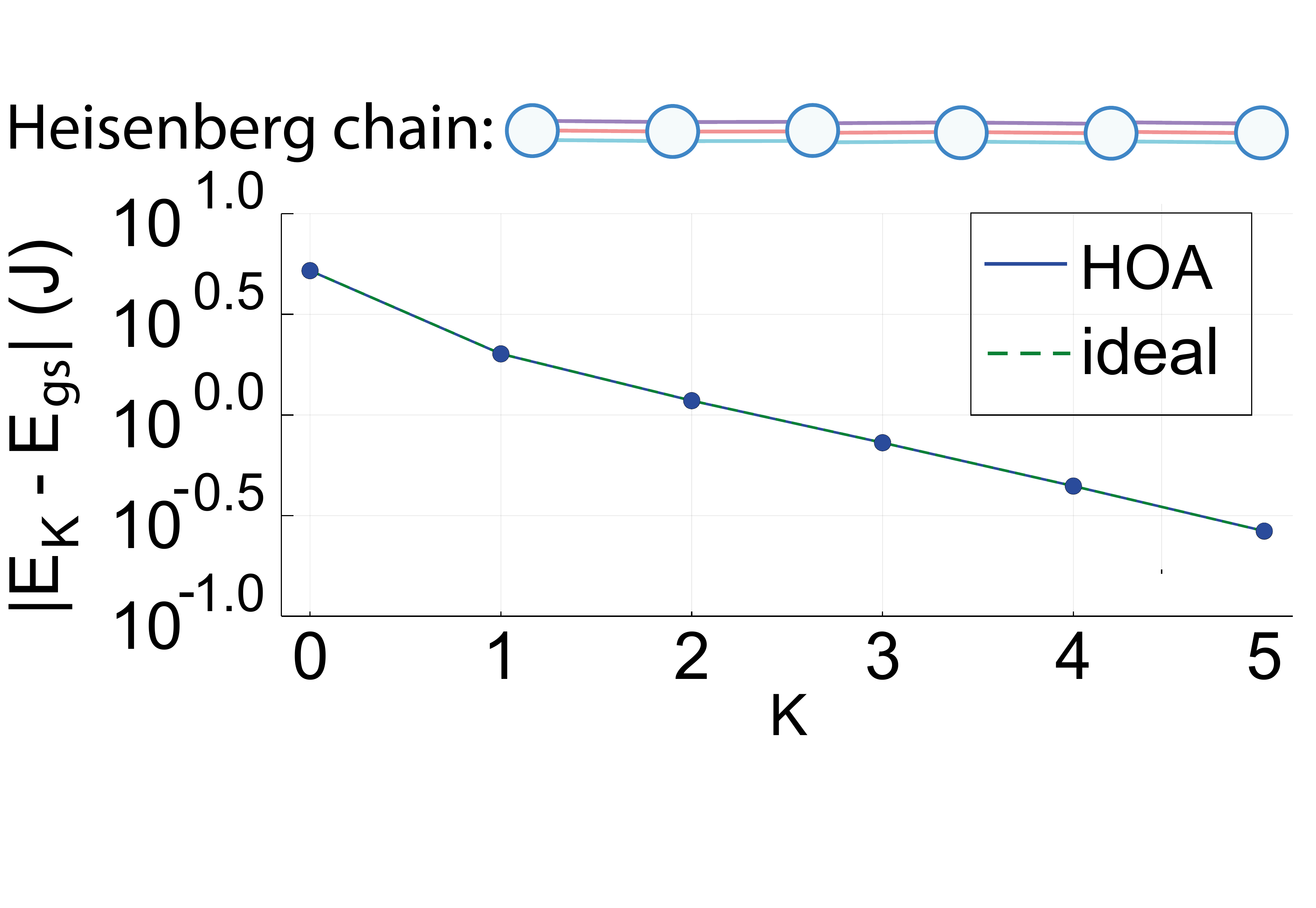}
\caption{Difference between the true ground state energy and the energy from the direct iteration method, where energy measurement is performed for the ideal and approximated Hamiltonians (lines overlay). We consider a six-qubit Heisenberg Hamiltonian at the critical point $h/J=1$, and HOA parameters of $\delta t = 10^{-1.75} J^{-1}$, $S=5$. }
\label{fig:image_direct_iterations}
\end{figure}

\subsection{Direct Hamiltonian iteration}

We generalize our consideration to approximate higher powers of the Hamiltonian operator $\hat{\mathcal{H}}^k$. This can be defined through the $k$-th order derivative of the propagator, $d^k\hat{\mathcal{U}}/dt^k = (-i \hat{\mathcal{H}})^k\hat{\mathcal{U}}(t)$, and following the numerical differentiation at $S$ stencil points we express it as 
\begin{equation}
        \hat{\mathcal{H}}^k = \frac{1}{(- i \delta t)^k} \sum_{n=-s}^{S-s-1} q^{(k)}_{n}(s) e^{- i\hat{\mathcal{H}} (n \delta t) }  + O\left(\delta t^{S-k}\right).
\end{equation}
The first nontrivial and useful example corresponds to the approximation of $\hat{\mathcal{H}}^2$ operator, allowing for the straightforward measurement of the variance. This is often required for variational schemes, as together with the energy it may help to detect the convergence of the algorithm \cite{Kokail2019}. Hamiltonian averaging through Pauli string measurement has daunting scaling in this case. We note that to get the same order of accuracy as for the expansion of $\hat{\mathcal{H}}$, we just need to increase the number of stencil points by one. This holds when going beyond $k=2$, where each power increment requires at least one additional stencil point, and we generally prefer using an odd number of stencil points for the interval to be symmetric. However, the scaling becomes unfavorable for small $\delta t$, with error growing as $O(\delta t^{S-1})$, thus shifting the optimal $\delta t$ region.

We proceed using HOA as a part of the direct iteration algorithm to prepare the dominant eigenstate of the Hamiltonian matrix. We start from an initial state $\ket{\Psi_0}$ having a non-vanishing overlap with the ground state. We simulate the repeated application of $\hat{\mathcal{H}}$ such that at iteration step $k$ we get $\ket{\Psi_{k}} = \hat{\mathcal{H}} \ket{\Psi_{k-1}}/ \Vert\hat{\mathcal{H}} \ket{\Psi_{k-1}}\Vert$, where the denominator accounts for the state normalization. At sufficiently large $k \mapsto K$ we get $\ket{\Psi_{K}} = \hat{\mathcal{H}}^{K} \ket{\Psi_0} / \Vert \hat{\mathcal{H}}^{K} \ket{\Psi_0}\Vert \approx  \ket{\text{ground state}}$, and corresponding expected energy
\begin{equation}
    E_{\text{ground}} \approx \bra{\Psi_{K}}\hat{\mathcal{H}}\ket{\Psi_{K}} = \frac{\bra{\Psi_0}\hat{\mathcal{H}}^{2K+1}\ket{\Psi_0}}{\bra{\Psi_0} \hat{\mathcal{H}}^{2K}\ket{\Psi_0}}
\end{equation}
approaches the ground state energy. We apply the described procedure to prepare a low energy state for the Heisenberg model at the critical point $h/J = 1$ (Fig.~\ref{fig:image_direct_iterations}). 
Starting from the product state corresponding to a mean-field solution, we lower the energy by one order of magnitude using just four iterations. We also observe that when $h/J>1$ the convergence further improves. At the same time, we note that direct iteration typically is an unstable procedure with critical dependence on the condition number, and works best for diagonally dominant matrices.
\begin{figure}[t]
\includegraphics[width=1.0\linewidth]{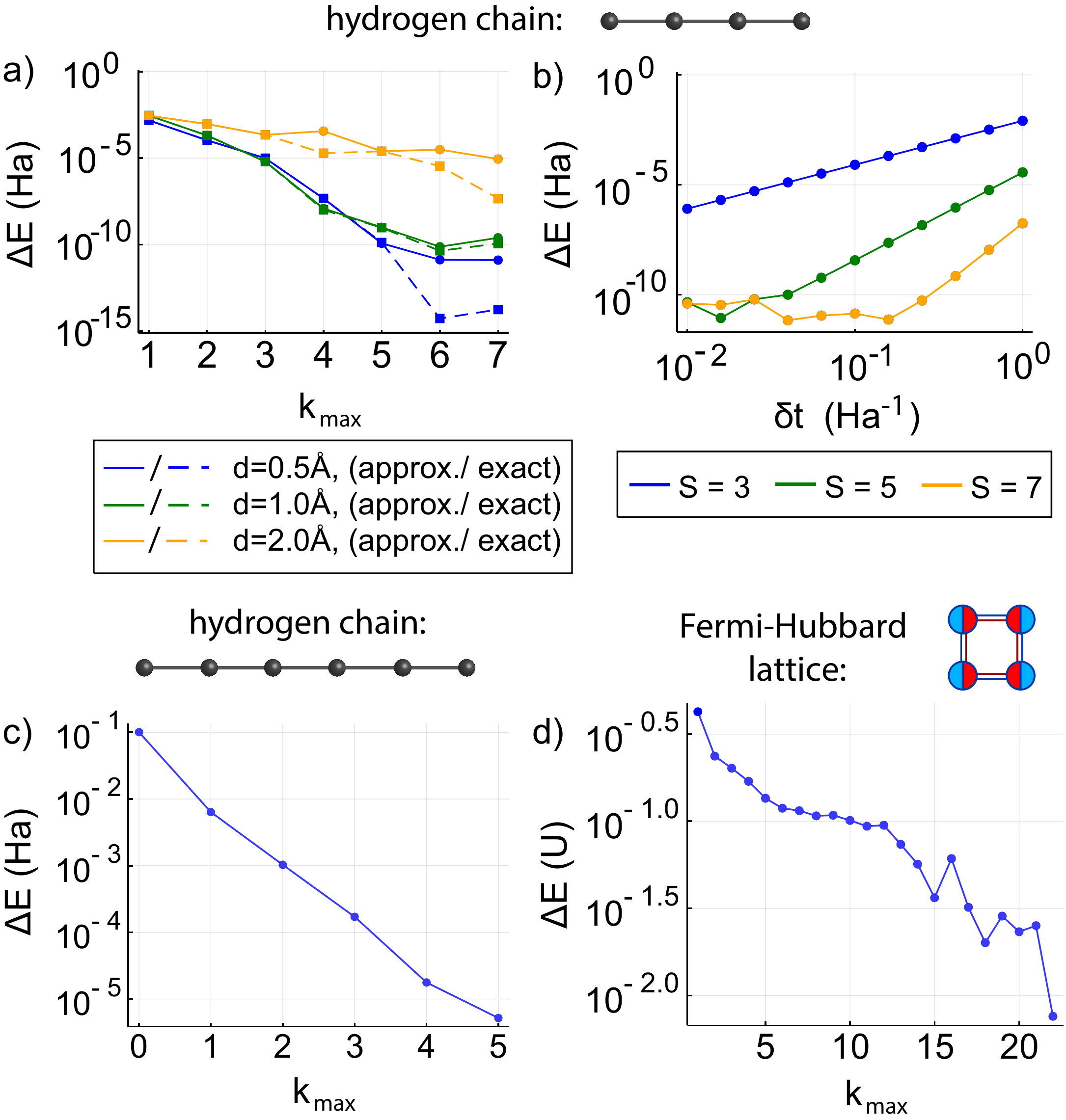}
\caption{a) Energy difference between the lowest state from the quantum filter diagonalization and true ground state, shown as a function of ansatz component number $k_{\mathrm{max}}$. Solid curves correspond to HOA, and dashed curves denote ideal Hamiltonian expectation. We consider molecular hydrogen H$_4$ at $d=0.5, 1.0, 2.0$~\AA, and use Gershgorin estimate $\kappa =  12.55, 5.93, 4.19$~Ha, with $S=5$, $\delta t = 0.01$~Ha$^{-1}$. b) Dependence of the HOA error on $\delta t$ for H$_4$ string at $d=1.0$~\AA, $k_{\mathrm{max}}=3$, $S=5$, and Gershgorin estimate $\kappa = 5.93$~Ha. c) Energy difference between QFD+HOA approach and true ground state as a function of $k_{\mathrm{max}}$. We consider H$_6$ molecule with $d=1.0$~\AA, S = 5, and Gershgorin estimate $\kappa = 12.86$~Ha. d) Energy difference between QFD+HOA and true ground state for four-site (8-qubit) 2D Fermi-Hubbard model with $J/U = 0.1$, $\mu /U = 0.05$, $h/U=0.001$, $\kappa/U = 4.60$, uniform initial state, $S=5$ stencils, and $\delta t U= 0.01$.}
\label{fig:H4_different_S}
\end{figure}

\subsection{Quantum filter diagonalization with Hamiltonian operator approximation}

Another application where our algorithm provides significant improvement is the quantum filter diagonalization (QFD) proposed by Parrish and McMahon \cite{Parrish2019}. The goal of the QFD procedure is to find the low energy spectrum of the system. Importantly, the procedure allows estimating both the ground state energy and excited state energies using the same resources (overlap measurements), thus being an efficient alternative to constrained variational methods \cite{Higgott2019,Nakanishi2019,Jones2019}. In QFD one uses the trial wavefunction as a sum of $2k_{\mathrm{max}}+1$ propagated states
\begin{equation}
    \ket{\Psi} = \sum_{j} \sum_{k=-k_{\text{max}}}^{k_{\text{max}}}c_{j,k} e^{-i \hat{\mathcal{H}} k/\kappa} \ket{\psi_j},
\end{equation}
where $\{ \ket{\psi_j} \}$ is a set of initial states and $\kappa$ is a spectral width parameter, which generally shall be greater than the difference of maximal and minimal eigenvalues $ |E_{\mathrm{max}} - E_{\mathrm{min}}|$. Next, the variational Rayleigh-Ritz approach is applied to find coefficients $c_{j,k}$ such that the energy functional $E = \bra{\Psi} \hat{\mathcal{H}} \ket{\Psi} /  \langle \Psi | \Psi \rangle$ is minimized. This corresponds to solving the generalized eigenvalue problem 
\begin{align}
\notag
    \sum_{j} \sum_{k=-k_{\text{max}}}^{k_{\text{max}}} c_{j,k} \bra{\psi_{j'}}e^{i \hat{\mathcal{H}} k'/\kappa} \hat{\mathcal{H}} e^{-i \hat{\mathcal{H}} k/\kappa} \ket{\psi_{j}} = \\ E \sum_{j} \sum_{k=-k_{\text{max}}}^{k_{\text{max}}} c_{j,k} \bra{\psi_{j'}} e^{i \hat{\mathcal{H}} k'/\kappa}e^{-i \hat{\mathcal{H}} k/\kappa} \ket{\psi_{j}}
\end{align}
where overlaps are calculated using a quantum circuit. Substituting $\hat{\mathcal{H}}$ with the differential approximation \eqref{eqExpansion} we avoid the Hamiltonian averaging procedure performed through Pauli string measurements, and the only difference for overlaps in LHS and RHS is the shift in evolution time by $(k - k')/\kappa$. As setting the time step for QFD requires the knowledge of the spectral width $\kappa$, in the following we use the Gershgorin circle theorem to provide its estimate in a scalable way \cite{Parrish2019}.

The results of QFD using Hamiltonian operator approximation are shown in Fig.~\ref{fig:H4_different_S}. First, we consider H$_4$ hydrogen chain ($N=8$ qubits) for different bond distances $d = 0.5$\AA, $1.0$\AA, $2.0$\AA~ and benchmark the difference between the true ground state and the variational state as a function of $k_{\mathrm{max}}$ (Fig.~\ref{fig:H4_different_S}a). We observe fast convergence to the true ground state, where starting from the Hartree-Fock initial state we can reach high precision ($<10^{-5}$ Ha) with only few components in the ansatz. We find that QFD results with the ideal Hamiltonian (dashed curves) and HOA (solid curves) only deviate at larger $k_{\mathrm{max}}$, and strong correlations at larger bond distance lead to a slower convergence rate. As $k_{\mathrm{max}}$ increases the expressibility of the ansatz improves, both due to larger number of states and longer propagation time. However, at large $k_{\mathrm{max}}$ the solution of generalized eigenvalue problem may be challenging due to instabilities, and care must be taken to choose it in the optimal way. In Fig.~\ref{fig:H4_different_S}b) we confirm that for fixed QFD parameters the quality of HOA improves with growing $S$ and decreasing $\delta t$. 
Going to the larger scale example of $N=12$-qubit H$_6$ hydrogen chain ($d=1.0$\AA), we again see exponential convergence to the groundstate energy with $k_{\mathrm{max}}$ increase, reaching chemical precision $\Delta E = 10^{-3}$~Ha already with five propagated states (Fig.~\ref{fig:H4_different_S}c).

Finally, we consider the challenging example of a spinful fermion lattice described by the Fermi-Hubbard model. This is described by the Hamiltonian that includes on-site Coulomb repulsion $U$ between opposite spins, nearest-neighbour hopping $J$, chemical potential $\mu$ (see the definitions and description in {\sffamily{}OpenFermion} package \cite{OpenFermion}). We consider a minimal two-dimensional lattice with four sites, and use the Jordan-Wigner transformation to write the $N=8$ Hamiltonian, where we additionally break the symmetry between up and down spin components with a weak effective magnetic field $h$. Choosing our initial state to be the uniform state, and staying at the half-filling, we observe that QFD+HOA approach can gradually bring the system towards the low energy state (Fig.~\ref{fig:H4_different_S}d). Importantly, HOA performs so well that no significant difference between approximation and the ideal Hamiltonian can be spotted, as the two curves overlay. Given the huge progress in analog quantum simulation of Fermi-Hubbard lattices with cold atoms~\cite{Mazurenko2017} and the possibility to perform interferometric measurements~\cite{Islam2015}, this suggests a promising route towards studying exotic fermion phases.

We note that the problem of preparing exact ground state of the many-body Hamiltonian is QMA-complete~\cite{Schuch2009}. This is defined by the overlap between initial state and the ground state, that may be exponentially small. However, the same concern holds for quantum phase estimation algorithms. Choosing physically motivated initial states that allow for efficient ground state preparation is an important goal for future studies in the field.

\section{Discussion}

In this section we discuss the prospects of using Hamiltonian operator approximation with near- and mid-term devices, and compare them to state-of-the-art techniques in the field.
\begin{figure}[t]
\includegraphics[width=1.0\linewidth]{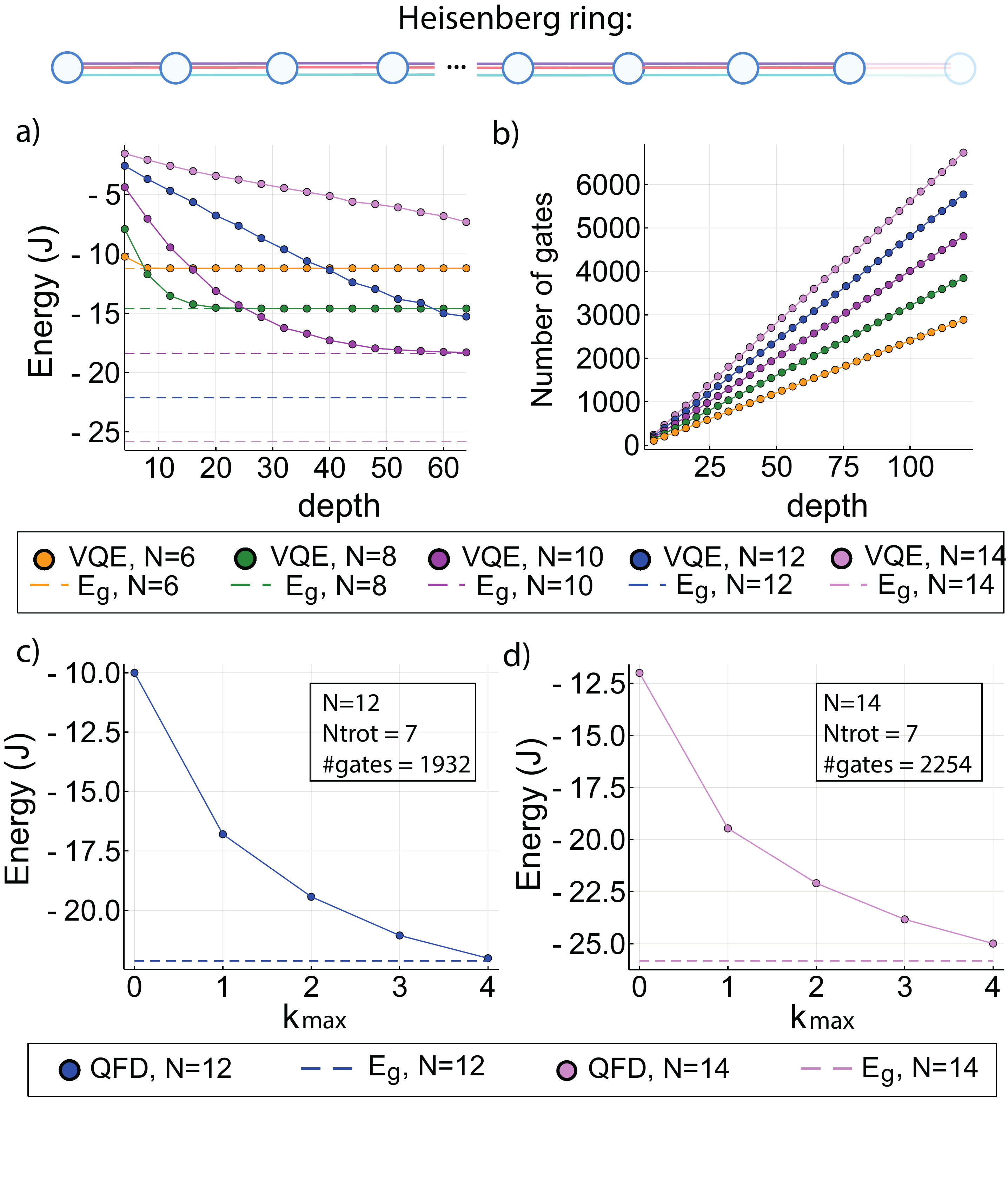}
\caption{a) Ground energy obtained from the VQE approach, shown as a function of the variational circuit depth. We consider spin-1/2 Heisenberg chain with 6, 8, 10, 12 and 14 qubits (ring geometry). Results are shown for $3000$ iterations of Adam optimizer, with analytical derivatives and the learning rate of $0.001$.  b) Total number of gates in a single variational circuit shown for increasing circuit depth. c) and d) Ground state energy of the Heisenberg ring estimated by the QFD+HOA method. System of $12$ and $14$ qubits are considered as a function of dynamical basis set size $k_{\mathrm{max}}$. We use a second-order Trotterization with $r=7$ Trotter steps.
}
\label{fig:VQE-comparison}
\end{figure}

\textit{VQE comparison.} 
First, we compare the QFD+HEA approach to the variational quantum eigensolver, being the standard NISQ tool. For this task we choose the problem of the ground state energy search for the Heisenberg ring at increasing system size. Specifically, we consider an additional magnetic field, such that the system is at the critical point ($h/J=1$) with highly entangled eigenstates. For the VQE we use the state-of-the-art optimization tool-set relying on analytic gradients (automatic differentiation) \cite{Mitarai2018,Schuld2019} and adaptive gradient descent \cite{Kingma2017}. It is known that automatic differentiation helps to mitigate the noise as compared to numerical differentiation techniques for updating the variational parameters~\cite{Kubler2020}. Specifically we used the Adam optimizer \cite{Kingma2017} widely used in machine learning, and enables the efficient search for optimal variational angles.

We also note that unlike chemistry problems that have a possibility to use the unitary coupled cluster ansatz~\cite{OxfordRev}, material science problems typically rely on the hardware efficient strategies, as employed for Heisenberg model in \cite{Kandala2017,Grant2019}. This corresponds to the ansatz choice as layers of generic rotations (concatenated parametrized rotations $\hat{R}_Z$-$\hat{R}_X$-$\hat{R}_Z$), followed by the network of CNOTs. Another possibility is to employ alternating operator ansatz, and specifically the Hamiltonian variational ansatz (HVA) with individual tunable blocks based on separate Hamiltonian terms \cite{Wiersema2020}. This can be seen as a physically motivated ansatz for spin systems.

The results for the statevector VQE simulation are shown in Fig.~\ref{fig:VQE-comparison}(a,b). We consider Heisenberg rings with the number of spins ranging from $6$ to $14$, and use the hardware efficient ansatz \cite{Kandala2017,Grant2019}. We optimize variational angles for maximum $3000$ iterations, setting Adam's learning rate to the  highest-performing value of $0.001$. We observe that for smaller system sizes ($N = 6, 8$) VQE can prepare the state being close to the true ground state, also staying within relatively shallow depth [Fig.~\ref{fig:VQE-comparison}(a)], and correspondingly small budget [Fig.~\ref{fig:VQE-comparison}(b)]. As $N$ grows the convergence to ground state requires ever increasing depth, and the corresponding gate count for reaching an approximate ground state at $N=14$ surpasses $6000$ gates (depth of $\sim 100$). While we do not observe yet the zero gradients regime, the ground state preparation requires a largely overparametrized circuit. Previously VQE with the depth of $73$ VQE was simulated for the same Heisenberg model at criticality for $N=7$ qubits \cite{Grant2019}. We perform additional tests with HVA (see details in Appendix \ref{C}), and observe similar convergence.

Now, let us check the cost of the quantum time grid approach, corresponding to HOA+VQE, for the same system. We use the Trotterization approach detailed in Appendix \ref{B}.
We find that for small systems the gate count from Trotterization dominates the budget of QFD+HOA, making it more costly than VQE for $N$ smaller than $10$ qubits. However, at $N=12$ it converges to the ground state with just $9$ propagated states and below $2000$ gates. For $N=14$ the convergence holds, needing $2250$ gates for the evolution. The weak scaling suggests that our approach has a larger constant as a starting budget, but outperforms VQE as the system grows to relevant sizes. We also note that our approach does not rely on optimization, and naturally avoids barren plateaus problem expected as the system scales to $\sim 20$ qubit size \cite{McClean2018}. These factors make the proposed approach suitable for NLSQ devices, while the point where it becomes beneficial depends on the task and hardware properties.

Comparing the measurement cost we see that for increased system sizes VQE needs over $3000$ iterations for convergence. Taking $N=14$ as an example, for each iteration the gradient measurement requires the number of independent runs being equal to six times the number of variational parameters ($\sim 6 \times 2000$), times the number of measurement shots (needs to be $>10^5$ for Adam, as shown in Ref. \cite{Kubler2020}). This leads to the measurement budget of $3\times 10^{12}$ shots. Our approach requires in general $N_{\mathrm{overlaps}} = (4 k_{\mathrm{max}} +1) S$ overlaps, being the number of time points of the QFD method multiplied by the number of points in the HOA method. This number decreases further if we adjust the time step of the HOA method to be equal to the time step in the QFD method. In such case many overlaps become degenerate, and the procedure only needs the calculation of $(4 k_{\mathrm{max}} +1) + 4$ unique overlaps. For $N=14$ we calculate $21$ unique overlaps, meaning that for the same measurement budget $10^{11}$ shots can be used to reach extra-high precision on each run. Conducting additional studies to compare the Hamiltonian averaging with HOA, we note similar $\propto 1/N_{\mathrm{meas}}$ drop in the variance for observables and overlaps, with the latter requiring a constant overhead. The point at which the total measurement budget for HOA becomes favorable then depends on the number of variational parameters and number of VQE iterations, ultimately defined by the efficiency of the variational workflow.

\textit{QPEA comparison.} 
Next, we compare the proposed approach to quantum phase estimation following \cite{ZapataRev}. The goal of QPEA is to determine eigenvalues of a unitary operator $\hat{\mathcal{U}}$, when acting on the corresponding eigenstate. While the full QPEA circuit requires an extended ancillary register and quantum Fourier transform \cite{Aspuru-Guzik2005}, the closest QPEA variation to quantum time grid methods is represented by Kitaev's phase estimation \cite{Kitaev1996,Dobsicek2007}. This requires only a single ancillary qubit and uses the adaptive protocol. However, to estimate the eigenvalue it uses the controlled unitary operation for the evolution operator, unlike separate overlap measurements. While in general the implementation of controlled unitaries is complex \cite{Zhou2011}, for the simulation of dynamics we can rewrite it as a simulation of a modified Hamiltonian with the reduced locality. Taking Heisenberg model as an example, this requires simulation of an effective three-body terms involving the ancilla coupled to other spins. Using the Trotterized evolution, we can decompose each Trotter step into single qubit rotations, Hadamards, phase gates, and CNOT operations. For QPEA this translates to controlled two-qubit gates and Toffoli gates, adding a significant ($\sim 10$) constant overhead, which increases further if restricted connectivity is considered. Also, QPEA is not applicable to analog quantum simulators. The relation between approaches thus depends on the platform and available quantum resources.

\textit{Scaling.} Finally, we ask the question: can Hamiltonian operator approximation become a viable strategy for the task of energy estimation in near- and mid-term future where larger systems are available but remain noisy? The issue of performing efficient energy measurement has recently gained attention \cite{Verteletskyi2020,Gokhale2019,Huggins2019b,Crawford2019,Cotler2020,Bonet2020}, and the advances are nicely summarized in Ref.~\cite{Bonet2020}, Table~I. In particular, considering vanilla Hamiltonian averaging with commuting Pauli heuristics one gets a simple measurement circuit (constant depth of Pauli rotations), but pays for it with $O(N^4)$ scaling for the number of partitions \cite{McClean2016} (we call it depth-frugal methods). On the other side, the methods based on the basis rotation grouping have much better scaling for the number of partitions being $O(N)$, while requiring measurement circuits with the gatecount of $N^2/4$ \cite{Huggins2019b} (we say they correspond to partition-frugal methods). Another related partition-frugal technique corresponds to the unitary partitioning approach proposed in \cite{Izmaylov2020}. Hamiltonian operator approximation thus takes the ultimate position in partition-frugal methods, where the number of independent terms (overlaps) to measure is minimized, at the expense of increased depth if a digital Hamiltonian simulation is used. The practical utility of HOA then depends on the maximal propagation time $T$ and the number of stencil points $S$ being used. Numerically we find that this number does not depend on the system size or has weak, at most $\log(N)$, dependence. The depth of the corresponding measurement circuit increases with $T$ translating to gatecount $\widetilde{O}(N^2 T)$ considering Trotterization \cite{Childs2019c}. However, the propagation time $T=S \delta t$ in HOA favors small values, and in many cases few Trotter steps suffice to reduce errors coming from the product formula (Appendix \ref{B}). Searching for examples where Hamiltonian operator approximation offers the advantage is an important task for the future research.

\section{Conclusions}

We proposed the Hamiltonian operator approximation technique that allows representing a Hamiltonian $\hat{\mathcal{H}}$ as a linear combination of unitary operators. Using numerical differentiation rules we rewrite $\hat{\mathcal{H}}$ as a sparse sum of quantum propagators, and benchmark it for relevant problems of energy estimation and ground state preparation. We found that the expected energy of the quantum system can be measured with high precision once we have access to the simulation of its dynamics, as for instance available in analog and digital quantum simulators. This holds in the presence of imperfections, including shot noise and decoherence. We also showed how the Hamiltonian operator approximation incorporates naturally in quantum Krylov-type approaches, and prepared the ground state of $12$-qubit H$_6$ hydrogen chain using the quantum filter diagonalization. Comparing the proposed approach to variational quantum eigensolver and Hamiltonian averaging we see it can become beneficial both in terms of gatecount and total number of shots for the increasing system size. However, the cross-over point depends on the task and quantum platform.

\textit{Note added.---}During completion of this work, we became aware of the independent work \cite{Seki2020} that has been carried out in parallel.

\begin{acknowledgments}
We thank Annie Paine and Salvatore Chiavazzo for useful discussions on the subject and reading the manuscript. T.A.B. thanks University of Exeter for hosting during the visit as a part of MSc project work. The authors acknowledge the support from the mega-grant No. 14.Y26.31.0015 of the Ministry of Education and Science of the Russian Federation and ITMO Fellowship and Professorship Program.
\end{acknowledgments}


\appendix
\section{Approximation errors}
\label{A}

Hamiltonian operator approximation relies on differentiation of the evolution operator, and introduces errors that depend on the finite differencing procedure. Our goal is to bound an approximation error as $\Vert \hat{\mathcal{H}} - \hat{H}_{\mathrm{diff}}(S,\delta t) \Vert < \epsilon$, where $\epsilon$ is a pre-defined constant. 
This shall be achieved for a minimal product of the step size and the total number of points in the differentiation grid, $S \delta t$.
Additionally, we note that small $S$ expansion is generally favoured due to its simplicity, and larger $\delta t$ helps at the stage where physical implementation errors are introduced.
The approximation error is a function of $S$, $\delta t$, and the Hamiltonian structure. 
Using the expansion procedure discussed before, for the infinite-precision arithmetic case the truncation error reads
\begin{align}
\label{eq:scaling}
    \Vert \hat{\mathcal{H}} - \hat{H}_{\mathrm{diff}}(S,\delta t) \Vert < C_S (\delta t)^S \max_t \left\Vert \frac{d^S}{dt^S} e^{-it\hat{\mathcal{H}}} \right\Vert,
\end{align}
where $C_S$ is a coefficient depending on the expansion. This suggests that the approximation improves as $S$ increases, and small $\delta t$ is beneficial. Similarly, we can write the bound for an expectation value of the Hamiltonian operator, and introduce a required expected energy precision as $\bra{\psi} \hat{\mathcal{H}} \ket{\psi} - \bra{\psi} \hat{H}_{\mathrm{diff}}(S,\delta t) \ket{\psi} < \epsilon_{\mathrm{energy}}$.
We further note that the scaling for HOA is in fact more involved as we approach the limit $\delta t \rightarrow 0$ \cite{DiffErr} and finite precision arithmetic is used. In this case the round-off error becomes important, and difference estimation is limited by minimal tolerance $\epsilon^*_{\mathrm{energy}}/\Vert \hat{\mathcal{H}} \Vert$, typically of the order of $10^{-16}$. For instance, in the case of quadratic approximation of derivative the total scaling reads as $\Vert \hat{\mathcal{H}} - \hat{H}_{\mathrm{diff}}(1,\delta t) \Vert < \varepsilon_{\mathrm{appr}} + \varepsilon_{\mathrm{num}}$, where $\varepsilon_{\mathrm{appr}} = \frac{1}{6} (\delta t)^2 \max_t \Vert d^S(\exp[-it\hat{\mathcal{H}}])/dt^S \Vert$ is the approximation error as in Eq.~\eqref{eq:scaling}, and $\varepsilon_{\mathrm{num}} = (\epsilon^*/\delta t) \max_t \Vert \exp[-it\hat{\mathcal{H}}] \Vert$ comes from the finite precision. We find that at small $\delta t$ the finite-precision rounding error $\varepsilon_{\mathrm{num}}$ starts to dominate, and there exists $\delta t^*$ such that the sum of the two contributions is minimized. However, we note that this limit is physically infeasible when the full measurement procedure is considered (see the discussion in the main text).

\section{Trotterization errors}
\label{B}

We aim to understand the scaling of Hamiltonian operator approximation for systems with Trotterized simulation of dynamics at increasing system size. For this we compare errors coming from the differential representation and Trotterization.

Specifically, we choose to simulate quantum dynamics digitally using the second-order Trotterization approach \cite{Tran2020}, and taking the Heisenberg Hamiltonian as an example. The Hamiltonian evolution for time $\tau$ can be simulated as
\begin{align}
e^{-i \hat{\mathcal{H}} \tau} \approx \Big(\prod_{j=N}^1 e^{(i J \tau/2r) \hat{X}_{j}\hat{X}_{j+1}} \prod_{j=N}^1 e^{ (i J \tau/2r) \hat{Y}_{j}\hat{Y}_{j+1}} \\ \notag
\prod_{j=N}^1 e^{(i J \tau/2r) \hat{Z}_{j}\hat{Z}_{j+1}} \prod_{j=1}^N e^{(i h \tau/r) \hat{Z}_{j}}   \prod_{j=1}^N e^{ (i J \tau/2r) \hat{Z}_{j}\hat{Z}_{j+1}} \\ \notag
\prod_{j=1}^N e^{ (i J \tau/2r) \hat{Y}_{j}\hat{Y}_{j+1}}\prod_{j=1}^N e^{(i J \tau/2r) \hat{X}_{j}\hat{X}_{j+1}} \Big)^{r} ,
\end{align}
where equality is valid for $J \tau \ll 1$ and $r\gg~1$.
For the chains with the periodic boundary we set indices as $N+1 = 1$. 
To simulate the circuit it is convenient to use M{\o}lmer-S{\o}rensen gates $\mathrm{XX}_{j,j+1}(\varphi) = \exp(-i \varphi \hat{X}_j \hat{X}_{j+1})$~\cite{Sorensen1999,Molmer1999,Maslov2017} acting on qubits $j$ and $(j+1)$ for phase $\varphi$. M{\o}lmer-S{\o}rensen gates are native to the trapped ions platform, and can be implemented efficiently beyond nearest neighbor connectivity. To simulate Heisenberg terms $\mathrm{YY}_{j, j+1}$ we then conjugate $\mathrm{XX}_{j,j+1}$ with the pairs of phase gates $\mathrm{S}$ acting at qubits $j$ and $(j+1)$. Namely, $\mathrm{YY}_{j, j+1}(\varphi) = \mathrm{S}_j \mathrm{S}_{j+1} \mathrm{XX}_{j,j+1}(\varphi) \mathrm{S}_j^{\dagger} \mathrm{S}_{j+1}^{\dagger}$,
where $\mathrm{S} = \mathrm{diag}(1,i)$. Similarly, we get $\mathrm{ZZ}_{j, j+1}$ by conjugating $\mathrm{XX}_{j, j+1}$ with Hadamard gates $\mathrm{H}$ as $\mathrm{ZZ}_{j, j+1}(\varphi) = \mathrm{H}_j \mathrm{H}_{j+1} \mathrm{XX}_{j,j+1}(\varphi) \mathrm{H}_j \mathrm{H}_{j+1} $.
\begin{figure}[t!]
    \centering
    \includegraphics[width=\linewidth]{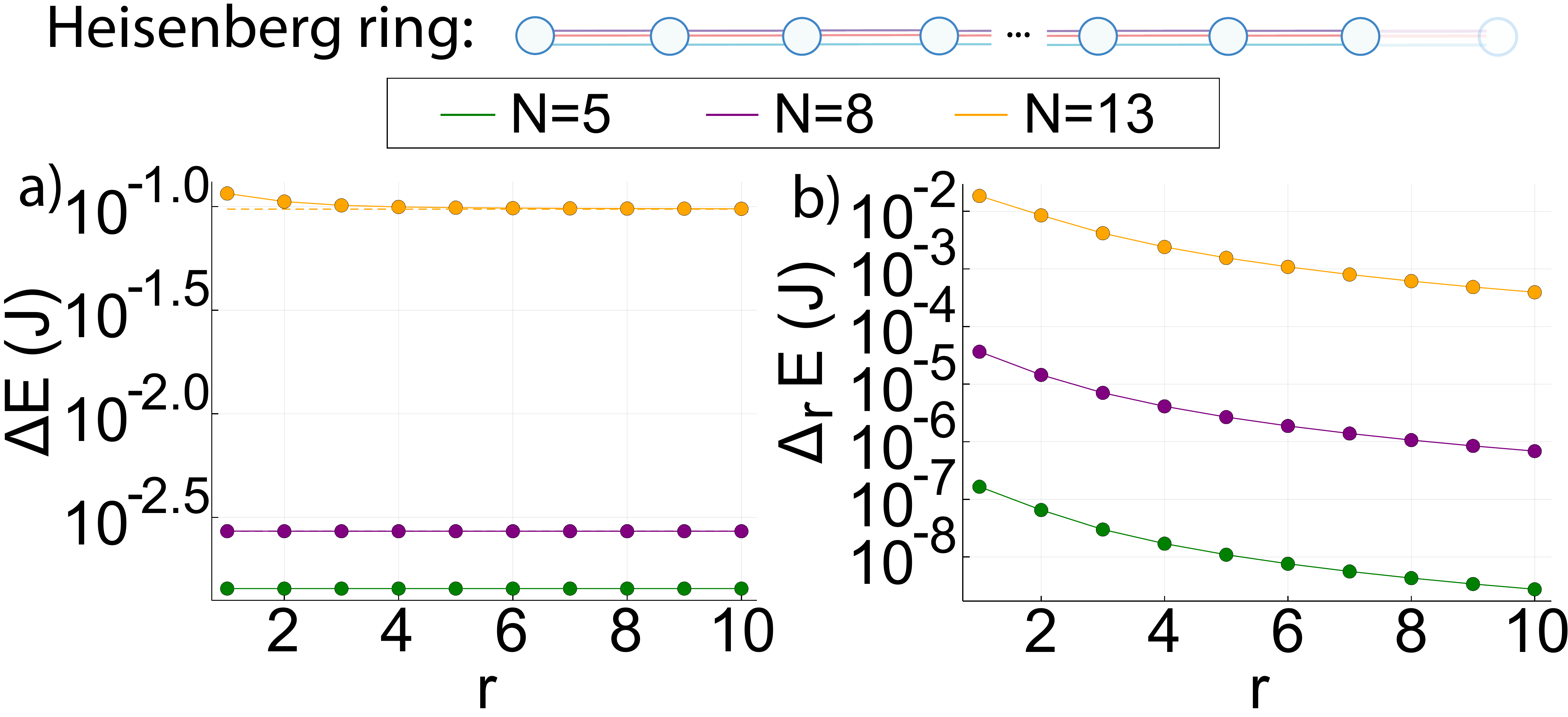}
    \caption{a) Difference between the energy obtained with the Trotterized HOA and exact energy of a state in the Heisenberg ring of $N=5$, $N=8$, $N=13$ qubits. Parameters of HOA approximation are $S=25$, $\delta t = 0.1$. b) Difference between the energy obtained with the Trotterized HOA and the energy from HOA with ideal evolution. The parameters are the same as in a). }
    \label{fig:TrotErr}
\end{figure}

In Fig.~\ref{fig:TrotErr} we show how the error caused by Trotterization (Fig.~\ref{fig:TrotErr}a) compares to the overall error of the HOA (Fig.~\ref{fig:TrotErr}b). We see that as the propagation times for HOA itself are relatively small it is only a small number of Trotter steps needed for the error caused by Trotterization to become much less than the HOA error. For $5$- and $8$-qubit Heisenberg rings the error caused by Trotterization is small already at $r = 1$. For a larger system of $13$ qubits with the same HOA parameters it is enough to have $5$ Trotter steps for the Trotter error to become negligible. In the simulation for Fig.~\ref{fig:TrotErr} we use values $\delta t J = 0.1$ and $S=25$, but note that typically these can be smaller, leading to even more pronounced decrease of the Trotterization error.

When considering the energy measurement as a part of the ground state preparation process we see that the maximal evolution time increases. In QFD+HOA this is posed by larger $k_{\mathrm{max}}$ as the system grows. In this case we expect increasing importance of the Trotterization error, that translates to larger gatecounts.

\section{VQE with Hamiltonian variational ansatz}
\label{C}

\begin{figure}[t!]
    \centering
    \includegraphics[width=0.9\linewidth]{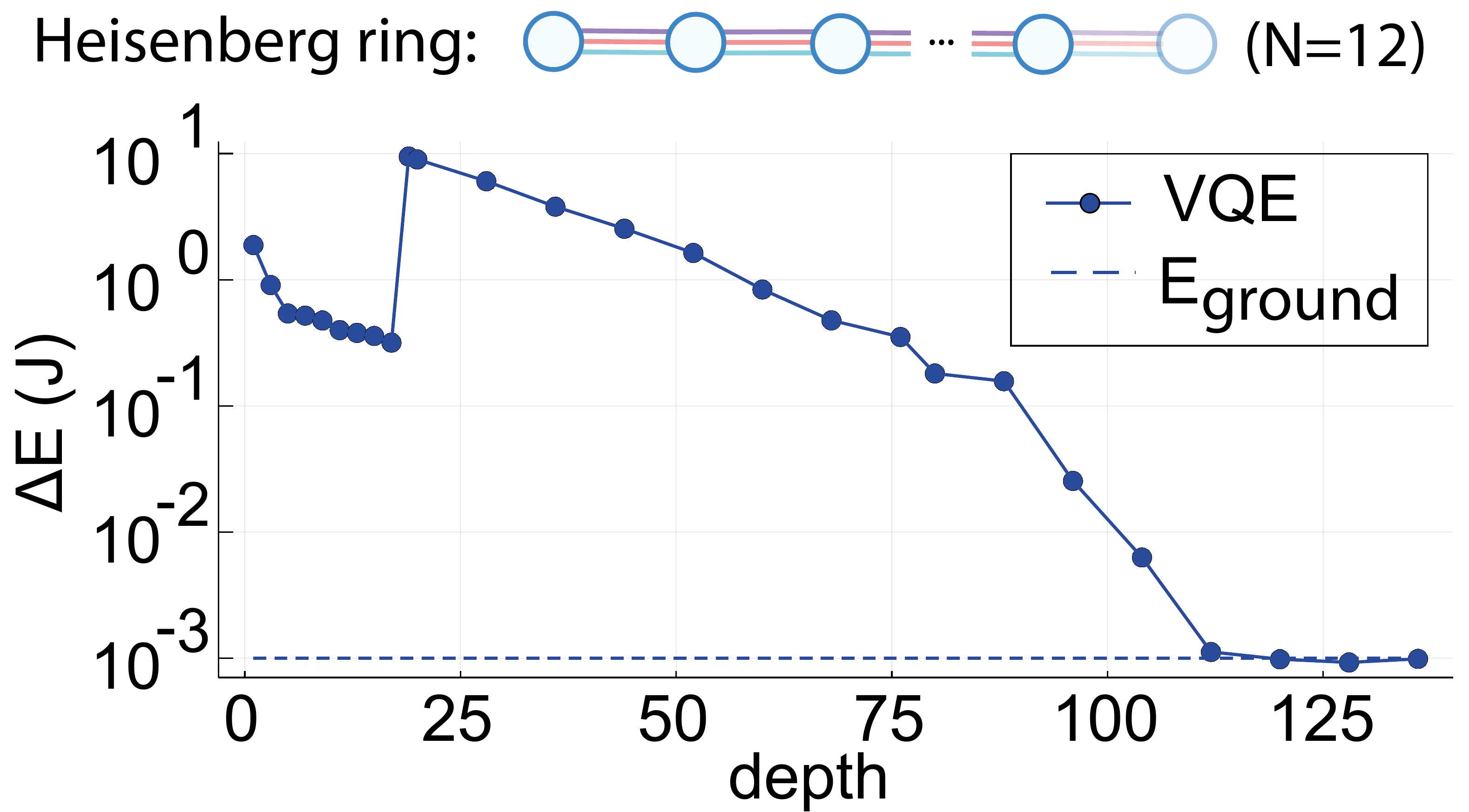}
    \caption{Energy difference of final VQE run and true ground state (log scale), shown as a function of HVA depth. The problem is the same as in Fig.~\ref{fig:VQE-comparison}, $N=12$. }
    \label{fig:HVA}
\end{figure}

In the main text for the VQE comparison we use the hardware-efficient ansatz based on layers of arbitrary single-qubit rotations and CNOTs. Additionally, we check the performance of the Hamiltonian variational ansatz for improving VQE convergence. The HVA circuit includes layers of Hamiltonian terms corresponding to $\hat{X}_{j} \hat{X}_{j+1}$, $\hat{Y}_{j} \hat{Y}_{j+1}$, $\hat{Z}_{j} \hat{Z}_{j+1}$, and $\hat{Z}_{j}$ evolution. Additional $\hat{R}_X$ and $\hat{R}_Z$ rotations are included to improve expressivity. We used stochastic gradient descent in the Adam form with the learning rate of $0.001$ and $3000$ iterations to achieve convergence. The results are shown and discussed in Fig.~\ref{fig:HVA}.
We observe the initial decrease of variational state energy as increasing depth ($d \sim N$), but note that circuits are not expressive enough to prepare the ground state with high fidelity. After the depth of twenty layers VQE reaches the barren plateaus, and halts the efficient search. With further increase of depth the overparametrization of ansatz improves the search, reaching high quality solution at large depth. We note that small gradients in this regime require increased number of shots to navigate the derivative-based optimization.



\end{document}